\documentclass[12pt,titlepage]{utarticle}
\usepackage{amsmath,amssymb,graphicx,color,amsthm}
\usepackage[curve]{xypic}
\usepackage{hyperref}

\newtheorem*{theorem*}{Theorem}

\numberwithin{equation}{section}

\def\BZ{\mathbb{Z}}

\newcommand{\MC}[1]{{\mathbb{C}^{#1}}}            
\newcommand{\PP}[1]{{\mathbb{P}^{#1}}}

\def\Spec{\mathrm{Spec}}
\def\Proj{\mathrm{Proj}}
\def\Hom{\mathrm{Hom}}

\def\Ext{\mathrm{Ext}}

\def\mcO{\mathcal{O}}

\def\ie{\textit{i.e.,\ }}

\begin{document}
\preprint{UTTG--09--05\\
\texttt{hep-th/0510158}\\}

\title{Moduli Spaces for D-branes at the Tip of a Cone}

\author{Aaron Bergman\address{Theory Group, Physics Department\\
             University of Texas at Austin\\
             Austin, TX 78712\\ {~}\\
             \email{abergman@physics.utexas.edu}}
             and Nicholas Proudfoot\address{Department of Mathematics\\
	   University of Texas at Austin\\
	   Austin, TX 78712\\ {~}\\
	   \email{njp@math.utexas.edu}}}
             
\Abstract{{\bf For physicists:}  We show that the quiver gauge theory derived from a Calabi-Yau cone
via an exceptional collection of line bundles on the base has the original cone as a component 
of its classical moduli space.  {\bf For mathematicians:}  We use data from the derived
category of sheaves on a Fano surface to construct a quiver, and show that its
moduli space of representations has a component which is isomorphic to the anticanonical
cone over the surface.}

\maketitle
\newpage

\section{Introduction}\label{sec:intro}

Building on the mathematical work \cite{BondalQuiv}, there has recently been tremendous progress 
in the use of exceptional collections to derive quiver gauge theories for Calabi-Yau cones 
\cite{Aspinwall:2004vm,BridgeTStruct,Herzog:2004qw, Aspinwall:2004mb,Bergman:2005ba,Aspinwall:2005ur}. 
In this program, initiated in \cite{Wijnholt:2002qz}, one begins with a 
Fano K\"ahler-Einstein surface $V$. 
An exceptional collection on $V$ is a collection of objects in the derived category 
of coherent sheaves on $V$ that forms an analogue of a basis of a vector space. 
Given such a collection with nice enough properties, Bondal \cite{BondalQuiv} constructs a quiver 
that has a derived category of representations equivalent to the derived category of coherent sheaves.  
It is shown in the above references how to `complete' the quiver to obtain a new,
more complicated quiver, whose category of representations corresponds not to $V$, but rather
to the total space $\omega$ of the canonical bundle over $V$.
In this paper we consider a related variety, $C(V)$, which (on the level of points) is
obtained from $\omega$ by collapsing the zero section.  It is so named because if the
anticanonical bundle of $V$ is very ample, then $C(V)$ is isomorphic to the cone
over $V$ in its anticanonical projective embedding.
It follows from topological string theory \cite{Aspinwall:2004mb} 
that the gauge theory on a D-brane located at the tip of the cone $C(V)$ 
is the quiver gauge theory corresponding to the completed quiver.
Other Calabi-Yau cones can be obtained by undoing an orbifold 
as shown in \cite{Bergman:2005ba}.  
Recently, Herzog and Karp \cite{Herzog:2005sy} have shown how to find exceptional 
collections describing a large class of toric cones, 
and Verlinde and Wijnholt \cite{Verlinde:2005jr} have applied these techniques towards 
string phenomenology.

In string theory, we generally expect that if we have a D-brane probing a particular geometry, 
the moduli space of the gauge theory on the brane (or perhaps a particular branch thereof) 
should correspond in some way to the geometry being probed. It then becomes an interesting 
question to ask about the moduli spaces of these quiver gauge theories derived 
from exceptional collections.
In mathematical language, this is the moduli space of representations of the 
\textit{completed} quiver. In this paper,
we will address this question for exceptional collections consisting solely of line bundles. 
(This implies that the ranks of the gauge groups are all one.) 
We will show that one component of the classical moduli space of the gauge theory is precisely 
the cone $C(V)$ on which the D-brane lives.

It is interesting to note that there exists a natural map from points on $V$ (respectively $\omega$) to the stack of isomorphism classes of representations of the original (respectively completed) quiver.\footnote{We thank David Ben-Zvi for calling our attention to this map.} A representation consists of a vector space for each node in the quiver and a linear map for each arrow.  In our case, the nodes of the quiver will correspond to the line bundles $E_1,\ldots,E_n$ in our
exceptional collection, and the set of arrows from $j$ to $i$ will correspond to vectors in the vector space $\operatorname{Hom}(E_i,E_j)$.  Given a point $p$ in our variety,
there is a canonical quiver representation with vector spaces given by the fibers at $p$
of the {\em dual} line bundles.
Set theoretically, the moduli space of the gauge theory is in a quotient of a subset
of the moduli stack.  The compatibility of the stack theoretic and moduli theoretic
approaches will be discussed in \cite{NickAaron}.

The main idea of our approach is as follows.
It is well-known in gauge theory that the moduli space of vacua is 
parameterized by the gauge-invariant operators
which for quiver gauge theories are all given by loops in the quiver.
(In mathematical language, the moduli space is equal to the
reduced variety underlying the affine GIT quotient of the space of representations
by the complexified gauge group, which is defined as $\Spec$ of the ring of gauge-invariant functions.)
A result of Bridgeland \cite{BridgeTStruct} tells us that the ring of based loops in our quiver,
\ie loops through a given node, is isomorphic to the ring of sections of the anticanonical line bundle on $V$.
It follows that the affine algebraic
variety parameterized by the ring of based loops is isomorphic to the cone $C(V)$.
There is still some difficulty arising from the mismatch between based loops and loops in general.
This is why $C(V)$ turns out to be only one irreducible component (branch) of the moduli space.

In this discussion, we have set the Fayet-Iliopoulis terms to zero.  (Mathematically, this means that we do not impose any stability condition on the space of representations.)  Turning on these terms corresponds to (partially) desingularizing the tip of the cone. The relation of this to dibaryonic operators in the $SU\!(d)^n$ gauge theory is currently being pursued by the first author.  We restrict in this paper to the situation where the cone is defined by the canonical line bundle. The extension to the undone orbifolds of \cite{Bergman:2005ba} is straightforward. It is also an interesting question, in the mathematical context, to ask whether the uncompleted quiver has a stability condition such that its moduli space of stable representations has
$V$ as one of its components. This will be addressed in a future work \cite{NickAaron}.

This paper is organized as follows. In Section \ref{sec:exceptional}, 
we briefly review the material that we need from the theory of exceptional collections
in the derived category of an algebraic variety.
In Section \ref{sec:moduli}, we 
describe quiver gauge theories and their moduli spaces and prove our main result. Finally, in Section \ref{sec:example}, we illustrate our result for the $\BZ_2$ orbifold of the conifold. 

\section{Exceptional collections and quivers} \label{sec:exceptional}

In this section, we give an overview of
the procedure for obtaining a quiver from an exceptional collection.
For more details, see 
\cite{BondalQuiv,Aspinwall:2004vm,BridgeTStruct,Herzog:2004qw,Bergman:2005ba}. 
An exceptional sheaf\footnote{This definition is most naturally applied to 
exceptional objects in the derived category, but in our case it is sufficient to work
with honest sheaves.} is a coherent sheaf, $E$, such that
\begin{equation}
\Ext^k(E,E) = \MC{\delta_{k0}}.
\end{equation}
A strong exceptional collection is an ordered collection of exceptional sheaves, $(E_1,\dots,E_n)$, such that
\begin{eqnarray}
&i>j \Rightarrow \Hom(E_i,E_j) = 0, &\\
&\text{and } \forall\, i,j,\quad\Ext^k(E_i,E_j) = 0 \quad \forall k \neq 0.&
\end{eqnarray}
Finally, the exceptional collection is called full if it generates the derived category. With this data, let
\begin{equation}
T = \bigoplus_{i=0}^n E_i\ ,
\end{equation}
and
\begin{equation}
A = \left[\Hom(T,T)\right]^\mathrm{op} = \bigoplus_{i\leq j}\Hom(E_i,E_j),
\end{equation}
where the multiplication
\begin{equation}\label{mult}
\Hom(E_i,E_j)\times\Hom(E_k,E_l)\to\Hom(E_k,E_j)
\end{equation}
is zero unless $i=l$ and is otherwise defined in the obvious manner.
Bondal \cite{BondalQuiv} constructs a quiver with relations whose nodes are indexed by the exceptional objects
$E_1,\ldots,E_n$, and whose path algebra is isomorphic to $A$.  In other words, the vector space of paths 
from node $j$ to node $i$ modulo relations is canonically isomorphic to $\Hom(E_i,E_j)$, and
the multiplication of Equation \eqref{mult} corresponds to composition of paths.
Furthermore, Bondal proves that the derived category of 
representations of this quiver is equivalent to the derived category of coherent sheaves on $V$.

Given such an exceptional collection, the doubly infinite collection of sheaves generated by the relation
\begin{equation}
E_i = E_{i+n} \otimes K_V, \quad i\in\BZ, 
\end{equation}
is called a helix.
Following Bridgeland \cite{BridgeTStruct}, we call this helix simple
if
$\Ext^k(E_i,E_j) = 0$ for all $i\le j \in \BZ$ whenever $k\neq 0$.
We next define the helix algebra,
\begin{equation}
\tilde{B} = \bigoplus_{k\ge 0} \prod_{j-i= k} \Hom(E_i,E_j),
\end{equation}
where $i$ and $j$ run over the integers. This algebra has a natural $\BZ$-action given by the isomorphism
\begin{equation}
\otimes K_V : \Hom(E_i,E_j) \longrightarrow (E_{i-n},E_{j-n}).
\end{equation}
The invariant subalgebra under this action is called the rolled-up helix algebra by Bridgeland; 
we will denote its opposite algebra by $B$.
It is has been long conjectured and finally shown in \cite{Aspinwall:2005ur} that this algebra 
is the path algebra of a `completed quiver' where the relations can be derived from a superpotential.

Recall that $\omega$ is defined to be the total space of the canonical bundle $K_V$ on $V$, and
let $\pi$ denote the projection from $\omega$ to $V$.
The nodes of the completed quiver correspond to the sheaves $\pi^*(E_i)$ on $\omega$, and
the arrows are defined such that the vector space spanned by the set of paths from node $j$ to node $i$
is isomorphic to
\begin{equation}
\label{pathloops}
\Hom(\pi^*(E_i),\pi^*(E_j)) = \bigoplus_{m\ge 0} \Hom(E_i,E_j \otimes K_V^{-m}).
\end{equation}
Note that these vector spaces are infinite dimensional, but each summand is finite dimensional.
The grading by $m$ in Equation \eqref{pathloops} corresponds to the grading in the path algebra given by
the number of times that a path circles the quiver.
Since we will be assuming that the sheaves $E_1,\ldots,E_n$ are all line bundles, Equation \eqref{pathloops}
tells us that for all $i$, the algebra $B_i$ of loops based at $i$ is isomorphic to
\begin{equation}\label{basedloops}
\Hom(\pi^*(E_i),\pi^*(E_i)) = \bigoplus_{m\ge 0} \Hom(K_V^{-m}).
\end{equation}
Even though the collection $\pi^*E_1,\ldots,\pi^*E_n$ on $\omega$ is not exceptional, Bridgeland shows that the
derived category of representations of $B$ and,
hence, of the completed quiver, is equivalent to the derived category of 
coherent sheaves on $\omega$.

\section{Moduli spaces of quiver gauge theories}\label{sec:moduli}

Given a quiver with relations derived from a superpotential, 
the additional ingredient that we need to define a quiver gauge theory is a dimension vector.
This is a vector of $n$ integers, $(d_1,\dots,d_n)$, where $n$ is the number of nodes in the quiver. The matter content then consists of a vector multiplet for each node associated with a $U(d_i)$ gauge group at that node and a chiral multiplet for each arrow that transforms in the fundamental of the head and the antifundamental of the tail.
We will restrict our attention to the case in which the gauge groups are all $U(1)$.
Thus each arrow corresponds to a complex number, and our configuration space is simply $\MC{\mathrm{\# arrows}}$. We then impose the F-term and D-term constraints with all the FI-terms set to zero, and quotient by the gauge groups. 
The F-terms are the relations in the quiver as discussed in Section \ref{sec:exceptional}, 
and they cut out a subvariety of $\MC{\mathrm{\# arrows}}$.  The D-term constraints 
are known in the mathematical literature as a moment map; imposing them 
and quotienting by the gauge group means taking a symplectic quotient.  
It is well known to both physicists and mathematicians \cite{Luty:1995sd,Kirwan} 
that this corresponds to taking a GIT quotient by the complexified gauge group. 
With the FI-terms turned off, this means that we consider the affine variety 
$\Spec\, R$, where $R$ is the ring of gauge invariant functions on our subvariety. 
In fact, the relevant space for physics is not $\Spec\, R$, but rather its underlying
reduced variety, \ie the set of points cut out of affine space by the equations defining the ring $R$.

Let us now consider the total space $\omega$ of the canonical bundle on $V$.
A polynomial function on $\omega$ may be thought of as an element of the section ring
\begin{equation}
S := \bigoplus_{m\ge 0} \Hom(K_V^{-m}),
\end{equation}
where $m$ corresponds to the degree of the polynomial in the fiber direction.
There is a natural projection from $\omega$ to $\Spec\, S$, and it is an isomorphism
away from the zero section, which gets collapsed to a point because $V$ is projective.
Let $C(V)$ be the underlying reduced variety of $\Spec\, S$.  If the anticanonical bundle
if very ample then $V$ is isomorphic to $\Proj\, S$, and $C(V)$ is simply the cone over $V$ in its
anticanonical embedding.
It is a theorem of LeBruyn and Procesi\footnote{See, for example, Lecture 10 of \cite{Derksen}.} that the invariant ring of any quiver with any dimension vector is generated by the traces of automorphisms 
coming from loops in the quiver.\footnote{It is interesting to note that if we replace the gauge groups by the special unitary groups $SU(d_i)$  we will have more invariants. In physics, these are called dibaryons, while in the mathematical literature, they are called semi-invariants of the quiver.}
Suppose that it is in fact generated by functions associated to loops that are based at a given node $i$.
Then Equation \eqref{basedloops} tells us that $R\cong B_i \cong S$, and therefore that 
$\Spec\, R\cong \Spec\, S$.  Passing to the underlying reduced varieties, we conclude that
the moduli space of vacua in the quiver gauge theory is isomorphic to $C(V)$.

Finally, we would like to eliminate the assumption that all invariants of the quiver
are generated by loops based at a single node.
In general, instead of an isomorphism between $R$ and $S$, we have the following 
commutative diagram for each $i$, where $\varphi_i$ is the isomorphism
of Equation \eqref{basedloops}. 
\begin{figure}[h]
\centerline{
\xymatrix{ 
B_i
\ar[dr]_{\sigma_i} 
&& S\ar[ll]_{\varphi_i} 
\\ 
& R \ar[ur]_{\Psi }}}
\end{figure}
The map $\Psi$ is determined by the property that the above diagram commutes for all $i$.  The existence of such a map is guaranteed by the
compatability of the isomorphisms $\varphi_i$, and its uniqueness
comes from the fact that the images of the various $\sigma_i$
generate $R$ as a ring.
The injection $\sigma_i\circ\varphi_i$ induces a surjective map from the quiver moduli space $\Spec\,R$
to $\Spec\,S$, while
$\Psi$ induces a section of this map.
We will prove below that there exists 
a nonzero elements $\beta_i\in B_i$ such that $\sigma_i$ becomes an isomorphism after inverting $\beta_i$.
It follows that the inclusion induced by $\Psi$ of $C(V)$ into the quiver moduli space 
is an isomorphism over the open set where $\beta_i$ is nonzero.
Since $C(V)$ is irreducible, this implies that the inclusion identifies $C(V)$
with an irreducible component of the moduli space.

Given any pair of nodes $i$ and $j$, Equation \eqref{pathloops} tells us that the space of 
paths in from $j$ to $i$ can be identified with the vector space
$\bigoplus_{m\ge 0} \Hom(E_i,E_j \otimes K_V^{-m})$.  Since $K_V^{-1}$ is ample,
this vector space is always nonzero.  Choose a pair of nonzero paths $p$ from $i$
to $j$ and $q$ from $j$ to $i$, and let $\beta_i^j$ be the composition $qp$.
If $\ell$ is a loop based at $j$, then
\begin{equation}\label{first}
\sigma_i(\beta_i^j)\cdot\sigma_j(\ell)
=\sigma_i(qp)\cdot\sigma_j(\ell)
=\sigma_i(p\ell q).
\end{equation}
Put $\beta_i = \prod_{j\neq i}\beta_i^j$.
Equation \eqref{first} tells us that $\sigma_i(\beta_i)\cdot\sigma_j(\ell)\in\sigma_i(B_i)$.
Since the ring $R$ is generated by functions associated to loops, this implies
that $\sigma_i(\beta_i)\cdot R$ is contained in $\sigma_i(B_i)$.  Thus, when we invert $\beta_i$,
the inclusion $\sigma_i$ becomes an isomorphism.

The precise mathematical theorem that we have proven may be stated as follows.

\begin{theorem*}
Let $E_1,\ldots,E_n$ be a full, strong, exceptional collection of line bundles
on a Fano surface $V$, generating a simple helix.  
Then $C(V)$ includes into the moduli space of S-equivalence
classes\footnote{This means that we identify two isomorphism classes if their
closures intersect in the moduli stack.}
of representations of the associated `completed quiver' with dimension vector
$(1,\ldots,1)$, and the image is the canonical reduced subscheme of an irreducible component.
\end{theorem*}

\section{An example}\label{sec:example}

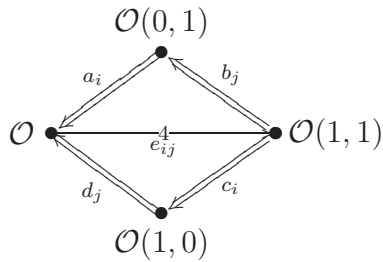
\begin{figure}[h]
\centerline{
\xymatrix{
{} & {\mathcal{O}(0,1)} \ar@2{<-}[]-/u 4mm/*{\bullet};[dr]-/r
8mm/*{\bullet}^{b_j} & {}\\
{\mathcal{O}} \ar@2{<-}[]-/l 4mm/*{\bullet};[ur]-/u 4mm/^{a_i} &
{} &
{\mathcal{O}(1,1)} \ar@{<-}[]-/r 8mm/;[ll]-/l 4mm/|4^{e_{ij}}
\ar@2{->}[]-/r 8mm/*{\bullet};[dl]-/d 4mm/*{\bullet}^{c_i} \\
{} & {\mathcal{O}(1,0)} \ar@2{->}[]-/d 4mm/;[ul]-/l 4mm/^{d_j} &
{}
}}
\caption{The completed quiver for $\PP{1} \times \PP{1}$.}
\label{p1p1cquiv}
\end{figure}

To illustrate our theorem, we consider an example in which 
$V \cong \PP{1}\times\PP{1}$, and $C(V)$ is the $\BZ_2$ orbifold of the famous conifold.  The line bundles $\mathcal{O}, \mathcal{O}(0,1), \mathcal{O}(1,0)$,
and $\mathcal{O}(1,1)$ form a full strong exeptional collection; its quiver is
shown in Figure \ref{p1p1cquiv}, with $i,j,k$ and $l$ running from 1 to 2.
The 4 in the middle of the central arrow means that there are four arrows from $\mcO$ to $\mcO(1,1)$.
Let $\{x_1,x_2\}$ be a basis of sections of $\mcO(0,1)$,
and $\{y_1,y_2\}$ a basis of sections of $\mcO(1,0)$. 
The arrows $a_i$ and $c_i$ correspond to multiplication by $x_i$,
and $b_j$ and $d_j$ correspond to multiplication by $y_j$. This gives the relations
\begin{equation}
\label{some}
a_ib_j = d_jc_i \,\,\text{  for all $i,j$}.
\end{equation}
In addition, the arrows $e_{ij}$ correspond to multiplication by $x_iy_j$ leading to further relations.
These relations can all be derived from the superpotential
\begin{equation}
W = (a_1b_1 - d_1c_1)e_{22} - (a_1b_2 - d_2c_1)e_{21} -
(a_2b_1 - d_1c_2)e_{12} + (a_2b_2 - d_2c_2)e_{11}.
\end{equation}

Let us consider only those loops that are based at the node $\mcO(0,1)$. 
There are na\"ively $2\times 4\times 2 = 16$ loops 
through this node that cycle exactly once around the quiver. 
Modulo relations, however, we find that the
ring of based loops is generated by nine monomials and may in fact be identified
with the subring of $\mathbb{C}[x_1,x_2,y_1,y_2]$ spanned by all monomials with the same even degree in both $x$ and $y$. One can see by explicit calculation that this gives the $\BZ_2$ orbifold of the conifold, but it is also possible to see this more geometrically. This ring is precisely the homogeneous coordinate ring of $\PP{1}\times\PP{1}$ in its projective embedding
defined by composing the $2$-uple embedding of $\PP{1}\times\PP{1}$ in $\PP{2}\times\PP{2}$ with the Segr\'e embedding
of $\PP{2}\times\PP{2}$ in $\PP{8}$. The hyperplane bundle on $\PP{8}$ pulls back to $\mcO(2,2) = K_V^{-1}$ over $\PP{1}\times \PP{1}$, so we immediately see that our variety is the affine cone we are looking for. In this case, the invariant ring is an integral domain, and, consequently,
any loop that does not pass through the node $\mcO(0,1)$ is equivalent in 
the path algebra to a linear combination of loops that do.

\section*{Acknowledgements}

Both authors would like to thank David Ben-Zvi, Jacques Distler, Sean Keel,
Deepak Khosla, and Uday Varadarajan for useful conversations.  For the first author, this material is based upon work supported by the National Science Foundation under Grant Nos. PHY-0071512 and PHY-0455649, and with grant support from the US Navy, Office of Naval Research, Grant Nos. N00014-03-1-0639 and N00014-04-1-0336, Quantum Optics Initiative.  The second author was partially
supported by an National Science Foundation Postdoctoral Research Fellowship.

\bibliographystyle{utphys}
\bibliography{thebib}

\providecommand{\href}[2]{#2}\begingroup\raggedright\begin{thebibliography}{10}


\bibitem{BondalQuiv}
A.~I. Bondal, ``Helices, representations of quivers and {K}oszul algebras,'' in
  {\em Helices and vector bundles}, vol.~148 of {\em London Math. Soc. Lecture
  Note Ser.}, pp.~75--95.
\newblock Cambridge Univ. Press, Cambridge, 1990.

\bibitem{Aspinwall:2004vm}
P.~S. Aspinwall and I.~V. Melnikov, ``D-branes on vanishing del Pezzo
  surfaces,'' {\em JHEP} {\bf 12} (2004) 042,
\href{http://www.arXiv.org/abs/hep-th/0405134}{{\tt hep-th/0405134}}.

\bibitem{BridgeTStruct}
T.~Bridgeland, ``{T-structures on some local Calabi-Yau varieties},''
  \href{http://www.arXiv.org/abs/math.AG/0502050}{{\tt math.AG/0502050}}.

\bibitem{Herzog:2004qw}
C.~P. Herzog, ``Seiberg duality is an exceptional mutation,'' {\em JHEP} {\bf
  08} (2004) 064,
\href{http://www.arXiv.org/abs/hep-th/0405118}{{\tt hep-th/0405118}}.

\bibitem{Aspinwall:2004mb}
P.~S. Aspinwall, ``D-branes, Pi-stability and Theta-stability,''
\href{http://www.arXiv.org/abs/hep-th/0407123}{{\tt hep-th/0407123}}.

\bibitem{Bergman:2005ba}
A.~Bergman, ``Undoing orbifold quivers,''
\href{http://www.arXiv.org/abs/hep-th/0502105}{{\tt hep-th/0502105}}.

\bibitem{Aspinwall:2005ur}
P.~S. Aspinwall and L.~M. Fidkowski, ``Superpotentials for quiver gauge
  theories,''
\href{http://www.arXiv.org/abs/hep-th/0506041}{{\tt hep-th/0506041}}.

\bibitem{Wijnholt:2002qz}
M.~Wijnholt, ``Large volume perspective on branes at singularities,'' {\em Adv.
  Theor. Math. Phys.} {\bf 7} (2004) 1117--1153,
\href{http://www.arXiv.org/abs/hep-th/0212021}{{\tt hep-th/0212021}}.

\bibitem{Herzog:2005sy}
C.~P. Herzog and R.~L. Karp, ``Exceptional collections and D-branes probing
  toric singularities,''
\href{http://www.arXiv.org/abs/hep-th/0507175}{{\tt hep-th/0507175}}.

\bibitem{Verlinde:2005jr}
H.~Verlinde and M.~Wijnholt, ``Building the standard model on a D3-brane,''
\href{http://www.arXiv.org/abs/hep-th/0508089}{{\tt hep-th/0508089}}.

\bibitem{NickAaron}
A.~Bergman and N.~Proudfoot, \textit{in preparation}.

\bibitem{Luty:1995sd}
M.~A. Luty and I.~Taylor, Washington, ``Varieties of vacua in classical
  supersymmetric gauge theories,'' {\em Phys. Rev.} {\bf D53} (1996)
  3399--3405,
\href{http://www.arXiv.org/abs/hep-th/9506098}{{\tt hep-th/9506098}}.

\bibitem{Kirwan}
F.~C. Kirwan, {\em Cohomology of quotients in symplectic and algebraic
  geometry}, vol.~31 of {\em Mathematical Notes}.
\newblock Princeton University Press, Princeton, NJ, 1984.

\bibitem{Derksen}
H.~Derksen, ``Lecture Notes on Quiver Representations,'', available at
\href{http://www.math.lsa.umich.edu/~hderksen/math711.w01/math711.html}{{\tt http://www.math.lsa.umich.edu/\~{}hderksen/math711.w01/math711.html}}\ .

\end{thebibliography}\endgroup
\end{document}